
\input polchar
\documentstyle{amsppt}
\NoBlackBoxes
\hoffset=.1in

\define\tr{\operatorname{Tr}}

\define\a{\alpha}
\define\be{\beta}

\define\gm{\gamma}
\define\de{\delta}
\define\si{\sigma}

\define\vd{\varDelta}
\define\vl{\varLambda}

\define\cb{\Bbb C}

\define\pd#1#2{\dfrac{\partial#1}{\partial#2}}

\define\vc#1{(#1_1,\ldots,#1_n)}
\define\vct#1{[#1_1,\ldots,#1_n]}
\define\vect#1{\{#1_1,\ldots,#1_n\}}


\define\jak{\frac{i}{\kappa}}

\define\frakap{\frac{1}{\kappa}}
\define\fraj{\frac{1}{2}}
\define\isl{ISL(2,\cb)}

\define\1{Poincare group}
\define\2{unitary}
\define\3{quantum}
\define\4{group}
\define\5{commut}
\define\6{representation}
\define\7{character}
\define\8{coequivariance}
\define\9{general}
\define\0{automorphism}

\document

\topmatter
\title The quantum $ISL(2,\cb)$ group
\endtitle
\rightheadtext{The quantum  group}
\author Pawe/l  Ma/slanka*\\
{\it Department of Functional Analysis}\\
{\it Institute of Mathematics, University of /L/od/x}\\
{\it ul. St. Banacha 22, 90--238 /L/od/x, Poland}\\
{e-mail  pmaslan\@ plunlo51.bitnet}
\endauthor
\leftheadtext{Pawe/l Ma/slanka}

\thanks
*\  Supported by KBN grant 2 0218 91 01
\endthanks
\abstract The classical $r$-matrix implied by the \3
$\kappa$-Poincare algebra of Lukierski, Nowicki and Ruegg is used
to generate a Poisson structure on the $\isl$ group (on the Hopf
algebra level) is obtained by a trivial quantization.
\endabstract

\endtopmatter
\document

Recently a \3 deformation of Poincare algebra has been
constructed which depends on dimensionful parameter $\kappa$ [1].
It has attractive properties like, for example, space isotropy or
the existence of natural cut-off $\kappa$. Some physical
consequences of deformed Poincare symmetry were considered by
Bacry [2]. The global counterpart of this algebra, i.e. the \3
Poincare group was constructed by Zakrzewski [3]. However, like
in the `classical' case for many purposes (the unitary \6s of
half-integer spin, supersymmetric extensions etc.) it is
important to consider the universal enveloping group
$ISL(2,\cb)$. It is the aim of the present letter to give a
description of this group. To this end we use the approach of
Zakrzewski and introduce the co-Poisson structure determined from
$\frac{1}{\kappa}$-expansion of the antisymmetric part of the
coproduct of the $\kappa$-Poincare algebra. By duality the
Poisson structure on $\isl$ is obtained and quantized. Due to the
form of Poisson brackets no quantization ambiguities appear.
Therefore one can expect that the `quasiclassical' (in $\frakap$)
duality can be extended to the full quantum duality (this is the
case for $E_q(2)$, (see  [4]).

The classical $\isl$ group consists of the pairs $(a,A)$ where
$a$ is a fourvector and $A$ is a matrix of the $SL(2,\cb)$ group
and with the composition law
$$
(a,A) * (a_1,A_1) = (a + \vl(A)a_1,AA_1)
\tag{1}
$$
where
$$
\vl(A) = [\vl^\mu_\nu(A)]^3_{\mu,\nu = 0} = [\fraj \tr (\si_\mu
A\si_\nu A^+)]^3_{\mu,\nu = 0}
\tag{2}
$$
is the Lorentz matrix corresponding to the matrix $A$ (and
$\si_\mu$ are the Pauli matrices).

In order to determine the Poisson structure on the $\isl$ group
we have to calculate the right and left-invariant vector fields.
To do this we can consider first the $GL(2,\cb)$ group. Due to
the fact that the $r$-matrix, which determines the Poisson
bracket by the formula
$$
\{f,g\} = 2r^{\a\be}(X^R_\a f X^R_\be g - X^L_\a f X^L_\be g)
\tag{3}
$$
where $ X^L_\a, X^L_\be$ are the right- and left-invariant vector
fields, contains only the generators of $SL(2,\cb)$ we can
consider all matrix elements $A^\a_\be$ as independent and impose
the unimodularity condition at the very end (i.e. $\det A$ has
vanishing Poisson bracket with everything). In this way we obtain
the following expressions for the invariant vector fields
$$
\aligned
(X^L)_\mu & = \vl^\nu_\mu(A)  \frac{\partial}{\partial a^\nu}\\
(X^L)^\a_\gm & = A^\be_\gm \frac{\partial}{\partial A^\be_\a}\\
(\bar X^L)^\a_\gm & = \bar A^\be_\gm \frac{\partial}{\partial
\bar A^\be_\a}\\
(X^R)_\mu & =  \frac{\partial}{\partial a^\mu}\\
(X^R)^\a_\be & = A^\a_\gm \frac{\partial}{\partial A^\be_\gm} +
\fraj (\si_\mu\si_\nu)^\a_\be a^\mu  \frac{\partial}{\partial a^\nu}\\
(X^R)^\a_\be & = \bar A^\a_\gm \frac{\partial}{\partial \bar A^\be_\gm} +
\fraj (\bar \si_\mu\bar \si_\nu)^\a_\be a^\mu
\frac{\partial}{\partial a^\nu}
\endaligned
\tag{4}
$$
$(\mu,\nu = 0,1,2,3$; $\a,\be = 1,2$).

In the Lie algebra basis corresponding to the above vector fields
the $r$-matrix obtained by Zakrzewski [3] takes the form
$$
r =  - \jak (L_k  \otimes  P_k - P_k   \otimes L_k)
\tag{5}
$$
where
$$
\align
L_k & = \frac{1}{2i} [(\si_k)^\a_\be X^\be_\a +
(\bar\si_k)^\a_\be \bar X_\a^\be]\\
P_k & = X_k
\endalign
$$
and $k = 1,2,3$.

This $r$-matrix induces (by duality) the following Poisson
brackets of the functions on $\isl$
$$
\aligned
\{f,g\} & = - \frac{1}{2\kappa}\{ [(\si_k)^\a_\be (X^R)^\be_\a f +
(\bar\si_k)^\a_\be ( \bar X^R)_\a^\be f] (X^R)_k g\\
& -[(\si_k)^\a_\be (X^R)^\be_\a g + (\bar\si_k)^\a_\be (\bar
X^R)_\a^\be g] (X^R)_k f\\
& -[(\si_k)^\a_\be (X^L)^\be_\a f + (\bar\si_k)^\a_\be (\bar
X^L)_\a^\be f] (X^L)_k g\\
& + [(\si_k)^\a_\be (X^L)^\be_\a g + (\bar\si_k)^\a_\be (\bar
X^L)_\a^\be g] (X^L)_k f\}.
\endaligned
\tag{6}
$$

If we perform the standard quantizations of the Poisson brackets
of the coordinate functions on $\isl$, by replacing $\{\, ,\, \}
\to \frac{1}{i}[\, ,\, ]$, one obtains the following set of
commutation relations
$$
\aligned
[A^\a_\be, A^\gm_\de] & = 0\\
[\bar A^\a_\be, \bar  A^\gm_\de] & = 0\\
[\bar A^\a_\be, A^\gm_\de] & = 0\\
[a^k,a^j] & = 0\\
[a^k,a^0] & = \jak a^k\\
[A^\a_\be, a^0] & = \frac{i}{2\kappa}(A\si_k)^\a_\be \vl^0_k(A)\\
[\bar A^\a_\be, a^k] & = \frac{i}{2\kappa}[(A\si_n)^\kappa_\be
\vl^k_n(A) - (\si_kA)^\a_\be]\\
[\bar A^\a_\be, a^0] & = \frac{i}{2\kappa}(\bar A\bar
\si_k)^\kappa_\be \vl^0_k(A)\\
[\bar A^\a_\be,a^k] & = \frac{i}{2\kappa}[(\bar A\bar
\si_k)^\kappa_\be \vl^k_n(A) -   (\bar \si_k\bar A)^\a_\be].
\endaligned
\tag{7}
$$
Since the composition law is compatible with Poisson bracket the
above \5ation rules are compatible with the following coproduct:
$$
\aligned
\vd(A^\a_\be) & = A^\a_\be  \otimes A^\gm_\be\\
\vd(a^\mu) & = \vl^\mu_\nu(A) \otimes a^\nu + a^\mu \otimes
I.
\endaligned
\tag{8}
$$
The antipode is given by
$$
S((a,A)) = (-\vl^\mu_\nu(A^{-1})a^\nu,A^{-1}).
\tag{9}
$$
We conclude that relations (7)-(9) define a Hopf algebra - $ISL_\kappa(2,\cb)$
group. Note that the map
$$
(a,A) \to (a,\vl(A))
\tag{10}
$$
is the Hopf algebra homomorphism from $ISL_\kappa(2,\cb)$ to the
$\kappa$-\1 of Zakrzewski.

{\bf Acknowledgment.}
\flushpar
I am grateful to  Prof. P. Kosi/nski for   discussions.

\enddocument